\documentclass[runningheads]{llncs}
\usepackage{graphicx}
\usepackage{float}
\usepackage{multirow}
\usepackage{caption}
\usepackage{subcaption}
\usepackage{booktabs}
\usepackage{varwidth}
\usepackage{cleveref}
\usepackage[table,xcdraw]{xcolor}
\begin{document}
\title{Don't Touch Me! A Comparison of Usability on Touch and Non-Touch Inputs}
\titlerunning{Don't Touch Me!}
%
\author{Kieran Waugh \and Judy Robertson}
\authorrunning{K. Waugh and J. Robertson}
%
\institute{The University of Edinburgh, Edinburgh, Scotland, UK \email{\{Kieran.Waugh,Judy.Robertson\}@ed.ac.uk}}
\maketitle              
\begin{abstract}
Public touchscreens are filthy and, regardless of how often they are cleaned, they pose a considerable risk in the transmission of bacteria and viruses. While we rely on their use, we should find a feasible alternative to touch devices. Non-touch interaction, via the use of mid-air gestures, has been previously labelled as not user friendly and unsuitable. However, previous works have extensively compared such interaction to precise mouse movements. In this paper, we investigate and compare the usability of an interface controlled via a touchscreen and a non-touch device. Participants (N=22) using a touchscreen and the Leap Motion Controller, performed tasks on a mock-up ticketing machine, later evaluating their experience using the System Usability and Gesture Usability scales. Results show that, in contrast to the previous works, the non-touch method was usable and quickly learnable. We conclude with recommendations for future work on making a non-touch interface more user-friendly.

\end{abstract}
\section{Introduction}
Public information kiosks and self-service displays are becoming common in public life, but there are some negative aspects to these types of public displays. Such devices have been found to have 1,475 times more bacteria than the average toilet seat~\cite{insuranceQuotes,NFS}. Similarly, the average supermarket self-service contains 5.9 times more bacteria than found on hospital displays~\cite{Gerba2016}. One possible solution to these issues is to remove the touch aspect of public devices and instead enable interaction through non-touch gestures. There is currently limited research on the relative usability of mid-air interactions in a public setting and few controlled studies that compare the Leap Motion Controller (LMC) and touchscreen for public interaction design. 
Our work explores the usability of non-touch technologies to replace public touchscreens. 

This paper documents a user study where participants compared and evaluated a mock-up ticket machine using both a touchscreen and a LMC which allows interaction via mid-air gestures. We measure participant success and errors when interacting with both devices as well as task time for the different inputs and system/gesture usability scores.

\section{Methodology}
We used a within-subjects study design where participants were asked to imagine they were at a train station. Tasks were then completed on both a touchscreen and LMC using on-screen point-and-push. To prevent step memorisation, the tasks in the two conditions differed slightly.
The order of the touch and non-touch conditions was randomised between participants and tasks were ordered so they increased in difficulty (Table \ref{tab:tasks}).
The system recorded an activity log for each participant and after completing each condition, participants were asked to fill out the System and Gesture Usability Scale surveys~\cite{brooke1996quick,Wickeroth2009} (SUS/GUS). After both conditions, they also filled out demographics, interface preference, and overall system feedback questions. Finally, we provided information about the cleanliness of the screen and asked for their preference again.
\vspace{-5mm}
\begin{table}[ht]
\caption{Tasks performed by the participants for both the touchcreen and LMC}
\centering

\resizebox{\textwidth}{!}{%
\begin{tabular}{|
>{\columncolor[HTML]{C0C0C0}}c |c|c|}
\hline
      & \cellcolor[HTML]{C0C0C0}Touch                                                                                                  & \cellcolor[HTML]{C0C0C0}Non-Touch                                                                                               \\ \hline
Task 1 & \begin{tabular}[c]{@{}c@{}}Buy a return ticket to Edinburgh Zoo \\ for two adults and one child\end{tabular}  & \begin{tabular}[c]{@{}c@{}}Buy a return ticket to Princes Street\\  for one adult and one child\end{tabular}    \\ \hline
Task 2 & \begin{tabular}[c]{@{}c@{}}Buy a one way ticket to The University \\ of Edinburgh for one adult and one child.\end{tabular} & \begin{tabular}[c]{@{}c@{}}Buy a one way ticket to Botanic Garden\\  for two adults and one child.\end{tabular} \\ \hline
Task 3 & \begin{tabular}[c]{@{}c@{}}Print a prepaid ticket using the code XBLPZ. \\ After, complete the survey.\end{tabular}            & \begin{tabular}[c]{@{}c@{}}Print a prepaid ticket using the code PRSGM. \\ After, complete the survey.\end{tabular}             \\ \hline
\end{tabular}
}
\vspace{2mm}

\label{tab:tasks}
\end{table}

\vspace{-8mm}

\begin{figure}[ht]

\centering
\begin{minipage}{.41\linewidth}
\centering
    \includegraphics[width=1\linewidth]{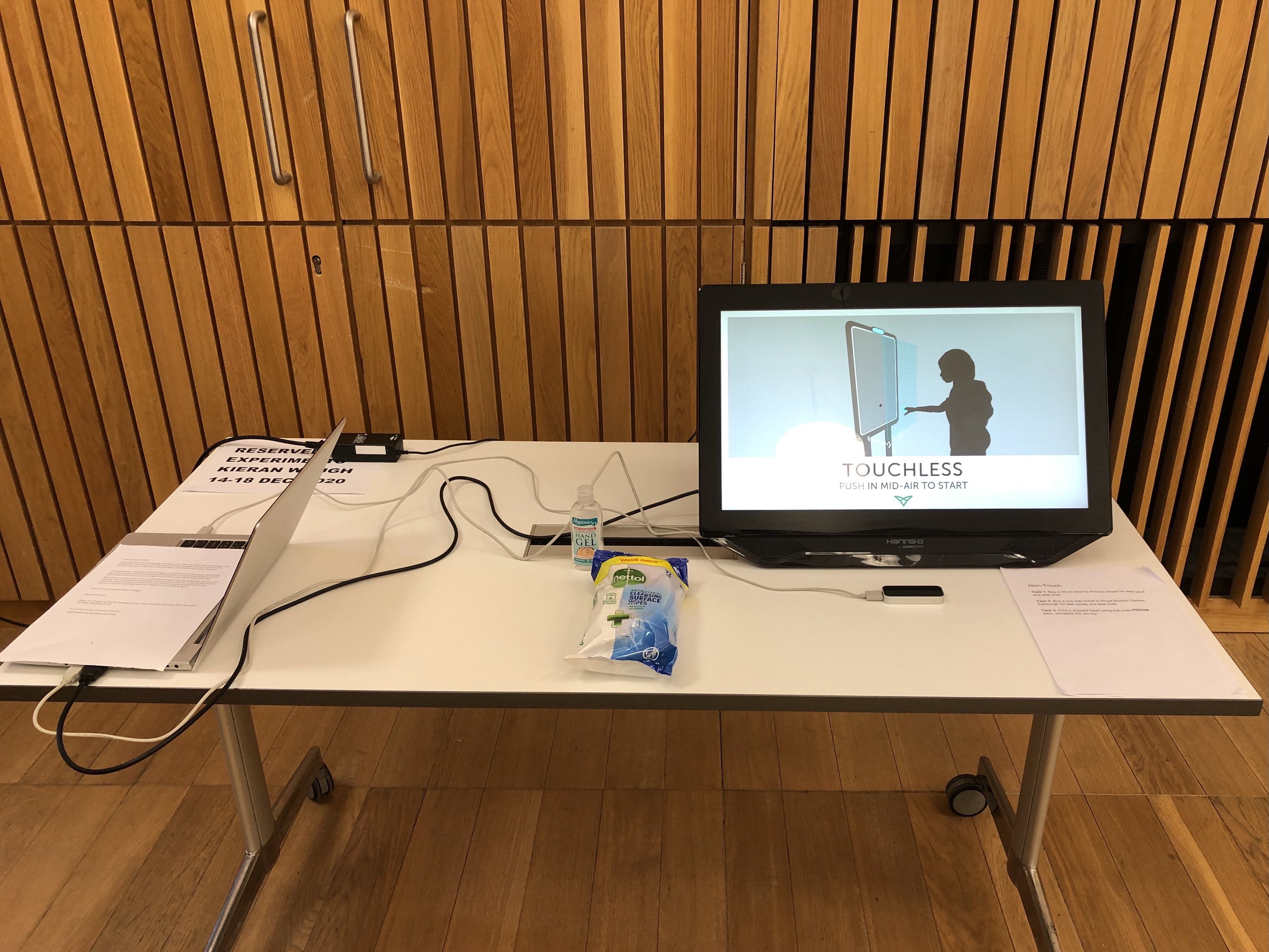}
    \caption{Study setup with touch monitor and LMC placed below.}
    \label{fig:setup}
\end{minipage}
\hfill
\begin{minipage}{.57\linewidth}
\centering
    \includegraphics[width=1\linewidth]{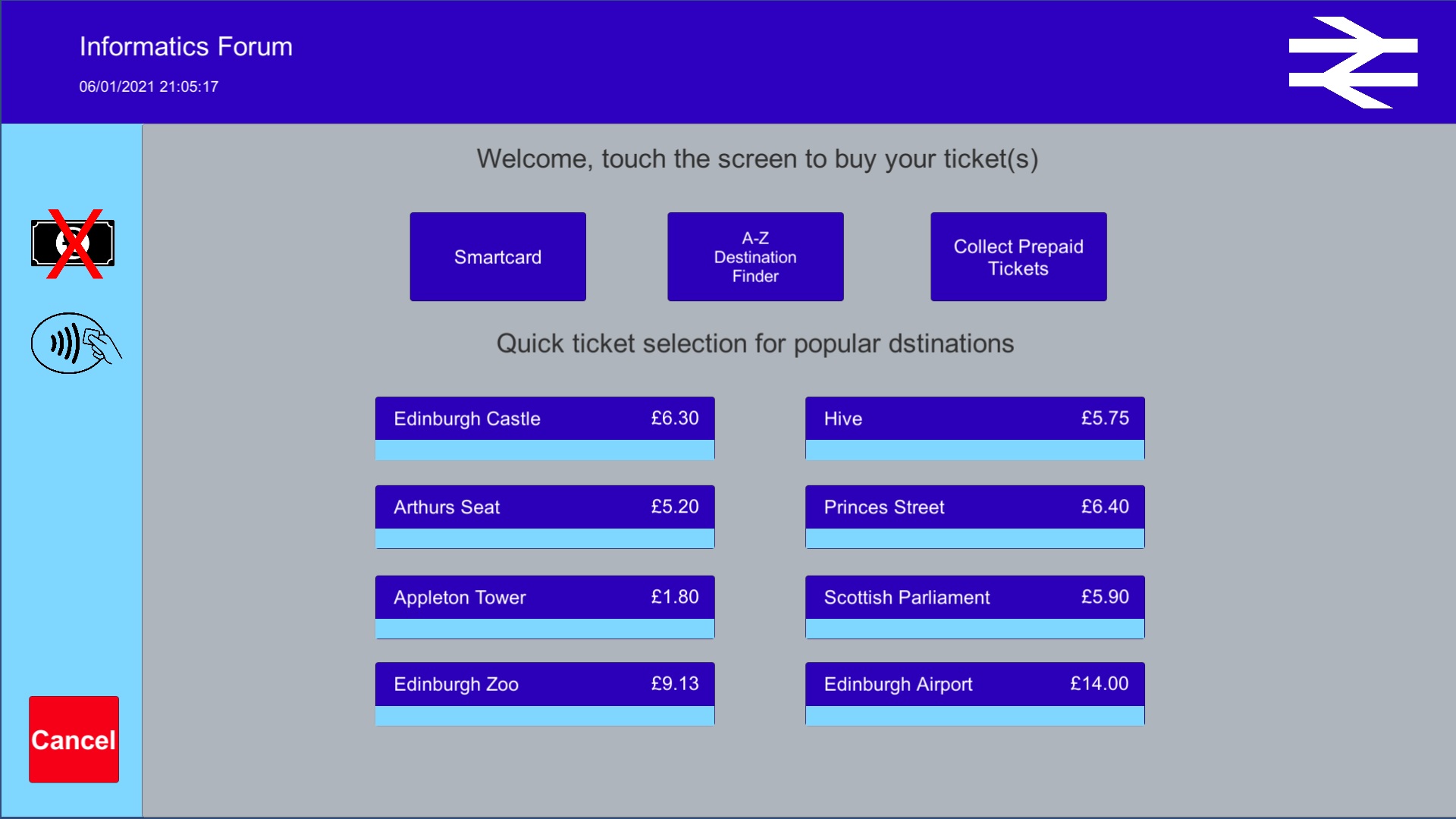}
    \caption{Interface shown to participants.}
    \label{fig:interface}
\end{minipage}
\end{figure}

\vspace{-10mm}
\section{Results}
22 people participated in the study, ranging from (self-reported) low to advanced technical backgrounds and an age range of 18 to 44. 8/22 participants preferred using non-touch over the touchscreen and 20/22 participants would use the system again in public. 12/14 people who preferred using the touchscreen changed preference after learning about the disease/bacteria spread on the average screen. Participants using the LMC took longer to complete the tasks (Table 2). Task 3, which required manipulating a UI slider, recorded the most time as the affordance suggested it should be draggable but a selection at a scale point was needed. On average, across the three tasks, participants made 2.09 errors with the LMC and 1.18 with the touchscreen with more errors made during task 2 and 3 (see Fig 3). 10/22 participants made 0 errors when using the LMC.

\vspace{-5mm}

\begin{table}[H]
\caption{Mean time in seconds and clicks per task with standard deviation (SD)}
\begin{center}
  \resizebox{0.65\textwidth}{!}{%
\begin{tabular}{|c|c|c|c|c|}
\hline
\rowcolor[HTML]{C0C0C0} 
\cellcolor[HTML]{C0C0C0}                   & \multicolumn{2}{c|}{\cellcolor[HTML]{C0C0C0}Mean Time (SD)} & \multicolumn{2}{c|}{\cellcolor[HTML]{C0C0C0}Mean Clicks (SD)} \\ \cline{2-5} 
\rowcolor[HTML]{EFEFEF} \hline
\multirow{-2}{*}{\cellcolor[HTML]{C0C0C0}} & Touch                        & Non-Touch                    & Touch                         & Non-Touch                     \\ \hline
\cellcolor[HTML]{C0C0C0}Task 1             & 23.9 (8.3)                 & 35.2 (11.5)                & 7.68 (1.39)                   & 6.91 (1.06)                   \\ \hline
\cellcolor[HTML]{C0C0C0}Task 2             & 32.0 (10.0)                & 53.9 (31.0)                & 7.73 (1.35)                   & 9.27 (1.98)                   \\ \hline
\cellcolor[HTML]{C0C0C0}Task 3             & 34.4 (8.1)                 & 64.8 (16.5)                & 11.09 (2.31)                  & 14.73 (2.69)                  \\ \hline
\end{tabular}%
}  
\end{center}

\label{tab:TimeClicks}
\end{table}

 \vspace{-10mm}
 As shown in Table 3, the touchscreen was rated higher in both scales, with a larger variance in the ratings for the non-touch scores. The post-experiment feedback was coded using open coding by the first author and later checked through discussion with the second author. Three categories were created: 1) user interface changes, 2) sensor tracking, and 3) general comments. 13 participants expressed a need to change the user interface (1), 11 identified issues with sensor tracking (2), and 3 participants provided further general comments (3). The participants identified specific points that could improve the system: clearer visual prompts, gesture pointers, bigger user interface elements and more space between them, and less use of the display corners. ``Bigger buttons may help too",``keyboard keys were a bit close together",``More difficult to select buttons in corners/edges could be improved".

     \setcounter{table}{3}
\vspace{-5mm}     
\begin{figure}[ht]
\centering
\begin{minipage}{.59\linewidth}
    \centering
    \includegraphics[width=1\linewidth]{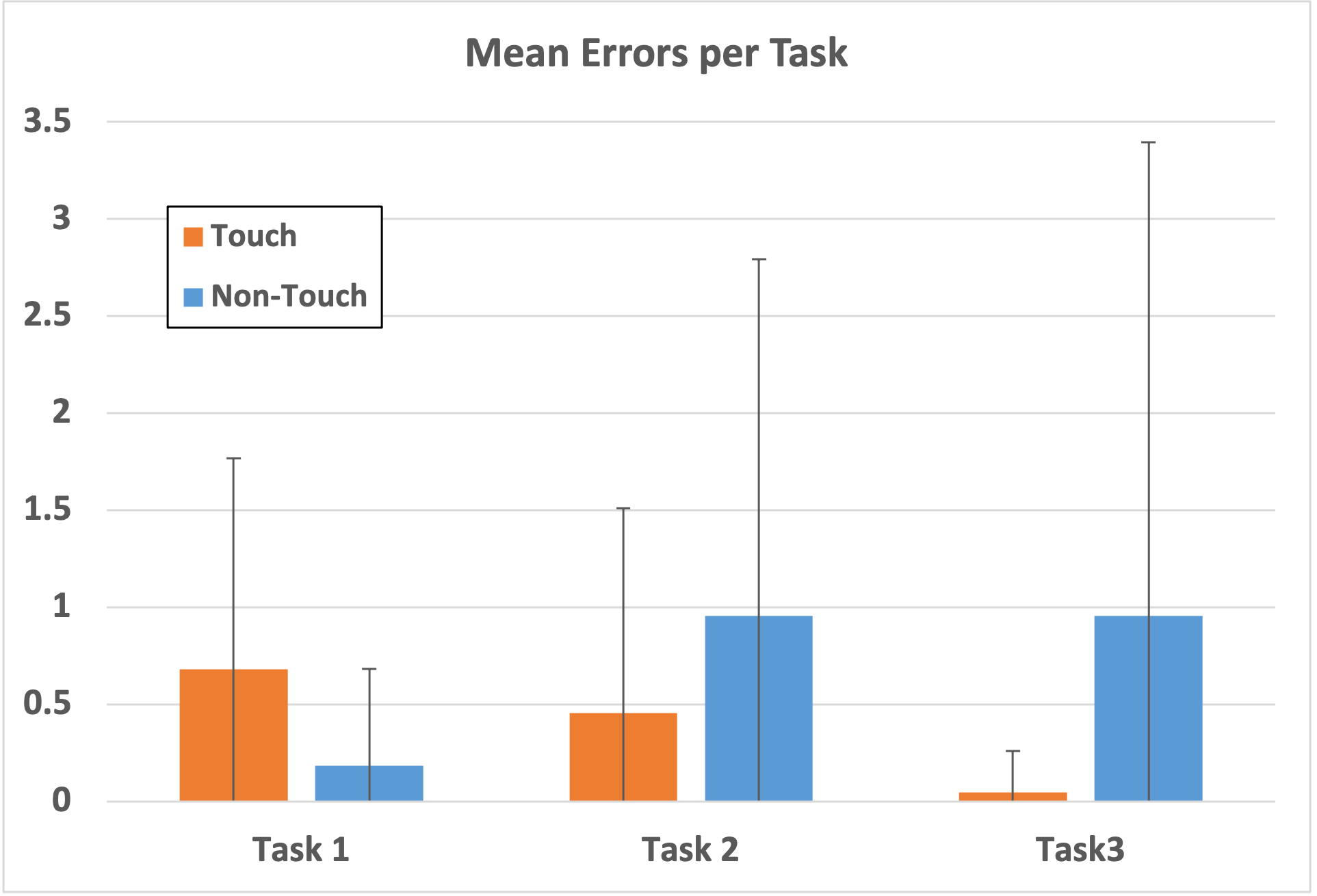}
    \caption{Mean errors per task with standard deviation}
    \label{img1}
\end{minipage}
\hfill
    \begin{minipage}{.4\linewidth}
    \centering
     \caption*{Table \thetable: Mean SUS and GUS scores for both Touch and Non-Touch with standard deviation}
        \begin{tabular}{|c|c|c|c|c|}
            \hline
            \rowcolor[HTML]{C0C0C0} 
            \cellcolor[HTML]{C0C0C0}                   & \multicolumn{2}{c|}{\cellcolor[HTML]{C0C0C0}Touch} & \multicolumn{2}{c|}{\cellcolor[HTML]{C0C0C0}Non-Touch} \\ \cline{2-5} 
            \hline
            \rowcolor[HTML]{EFEFEF} 
            \multirow{-2}{*}{\cellcolor[HTML]{C0C0C0}} & Mean                     & SD                      & Mean                      & SD                         \\ \hline
            \cellcolor[HTML]{C0C0C0}SUS                & 90.34                    & 10.33                   & 70                        & 18.39                      \\ \hline
            \cellcolor[HTML]{C0C0C0}GUS                & 95.3                     & 8.38                    & 64                        & 24.33                      \\ \hline
        \end{tabular}
    \end{minipage}
\end{figure}

 
 \vspace{-8mm}
 \section{Discussion}
 
Previous research comparing the LMC with traditional mouse actions found the LMC to have poor usability~\cite{SiexasLeap2D,BachmannLMEval,Saalfeld2015}. In contrast to these previous works, in our study the LMC shows promising usability with participants scoring their experience highly. As shown in Table \ref{tab:SusComp}, our SUS score is not only higher than the published average but also far higher than those in previous works. A possible explanation for these higher scores may be a more mature technology with better tracking. Having a high SUS/GUS score in this experiment along with a relatively low error rate suggests poor usability may no longer be the case.
 
The qualitative feedback indicates that design changes to the user interface could improve usability as also highlighted by Bachmann et al. \cite{BachmannLMEval}. Participants highlighted difficulties with the gesture set. Particularly, they were frustrated at the accuracy of point and push required by the LMC. These gestures are problematic as they attempts to follow touchscreen conventions, further supporting the idea of creating non-touch designed gesture sets \cite{Saalfeld2015}. There is a clear gap in research on user interfaces designed for public use with non-touch devices.

\vspace{-7mm}

\begin{table}[ht]
\caption{Comparison of previous studies of LMC and other input devices.}
\centering
\label{tab:SusComp}
\begin{tabular}{|
>{\columncolor[HTML]{C0C0C0}}c |c|c|c|}
\hline
                              & \cellcolor[HTML]{C0C0C0}SUS LMC & \cellcolor[HTML]{C0C0C0}SUS Other & \cellcolor[HTML]{C0C0C0}Device Compared \\ \hline
Sauro  \cite{Sauro2011}      & -                               & 68                                & Published average                       \\ \hline
Pirker et al. \cite{SUSLMC2}  & 55                              & 75                                & Keyboard                                 \\ \hline
Škrlj et al. \cite{Skrlj2014} & 56.5                            & 88.7                               & 3D Mouse                                    \\ \hline
Our study                     & 70                              & 90.3                             & Touchscreen                             \\ \hline
\end{tabular}
\end{table}
 
 \vspace{-8mm}
 \section{Conclusions and Future Work}

An exploration into touchless technology is an important first step to creating a feasible alternative to touch devices. Our results indicate that a non-touch approach is usable in a public setting and quickly learnable. While participants overall preferred the touchscreen, the resulting data for the LMC is significantly more promising than those in previous works, signalling considerable improvements in the gesture tracking technology. Further investigation is required with a more diverse user group to help build an accessible user interface for non-touch systems. Two areas need to be addressed in further research:
\begin{enumerate}
    \item A user interface specifically designed for a non-touch environment. Given the feedback from this study, potential areas of improvement are the size and location of UI buttons/objects and clearer calls to interact.
    \item Exploration of alternative interaction gestures. The gesture set chosen for this experiment did not work for all elements. The user interface must be designed with consideration to the limitations of the input range. For example, in this study, a UI slider not working with a point and push gesture.
\end{enumerate}
Further investigation into specialised user interfaces and gesture sets can advance the potential of non-touch systems, facilitating the adoption of this technology into public spaces.

\subsubsection*{Acknowledgements}
The authors would like to thank everyone who took time to participate in the study and Kate Farrell for supplying equipment. We would also like to thank Kami Vaniea for their advice and my colleagues and friends for their help proof-reading.

%
%
%
%
\bibliographystyle{splncs04}
\bibliography{references}
\end{document}